\documentclass[runningheads]{llncs}
\usepackage{graphicx}
\usepackage{authblk}
\usepackage{graphicx}
\usepackage{subfigure}
\usepackage{array}
\usepackage{float}
\usepackage{booktabs}
\usepackage{bbding}
\usepackage{etoolbox}
\usepackage{color}
\usepackage{soul}
\usepackage{algorithm}
\usepackage{algorithmic}
\usepackage{cite}
\usepackage{balance}
\usepackage{amsmath, amssymb}
\usepackage{makecell}
\usepackage{multirow}
\usepackage{hyperref}

\begin{document}
\title{LLCaps: Learning to Illuminate Low-Light Capsule Endoscopy with Curved Wavelet Attention and Reverse Diffusion}
\titlerunning{LLCaps: Learning to Illuminate Low-Light Capsule Endoscopy}

\authorrunning{L. Bai et al.}

\author{Long Bai\inst{1~\star} 
\and Tong Chen\inst{2} 
\thanks{Long Bai and Tong Chen are co-first authors.}
\and Yanan Wu\inst{1,3} 
\and An Wang\inst{1} 
\and Mobarakol Islam\inst{4} 
\and Hongliang Ren\inst{1,5} 
\thanks{Corresponding author.}}
\institute{Department of Electronic Engineering, The Chinese University of Hong Kong (CUHK), Hong Kong SAR, China
\and The University of Sydney, Sydney, NSW, Australia
\and Northeastern University, Shenyang, China
\and Wellcome/EPSRC Centre for Interventional and Surgical Sciences (WEISS), University College London, London, UK
\and Shun Hing Institute of Advanced Engineering, CUHK, Hong Kong SAR, China\\
\email{b.long@link.cuhk.edu.hk, tche2095@uni.sydney.edu.au, yananwu@cuhk.edu.hk, wa09@link.cuhk.edu.hk, mobarakol.islam@ucl.ac.uk, hlren@ee.cuhk.edu.hk}}
\maketitle              
\begin{abstract}
Wireless capsule endoscopy (WCE) is a painless and non-invasive diagnostic tool for gastrointestinal (GI) diseases. However, due to GI anatomical constraints and hardware manufacturing limitations, WCE vision signals may suffer from insufficient illumination, leading to a complicated screening and examination procedure. Deep learning-based low-light image enhancement (LLIE) in the medical field gradually attracts researchers. Given the exuberant development of the denoising diffusion probabilistic model (DDPM) in computer vision, we introduce a WCE LLIE framework based on the multi-scale convolutional neural network (CNN) and reverse diffusion process. The multi-scale design allows models to preserve high-resolution representation and context information from low-resolution, while the curved wavelet attention (CWA) block is proposed for high-frequency and local feature learning. Moreover, we combine the reverse diffusion procedure to optimize the shallow output further and generate images highly approximate to real ones. The proposed method is compared with eleven state-of-the-art (SOTA) LLIE methods and significantly outperforms quantitatively and qualitatively. The superior performance on GI disease segmentation further demonstrates the clinical potential of our proposed model. Our code is publicly accessible at
\href{https://github.com/longbai1006/LLCaps}{github.com/longbai1006/LLCaps}.

\end{abstract}
\section{Introduction}
\begin{figure*}[!t]
    \centering
    \includegraphics[width=1\linewidth, trim=25 375 295 0]{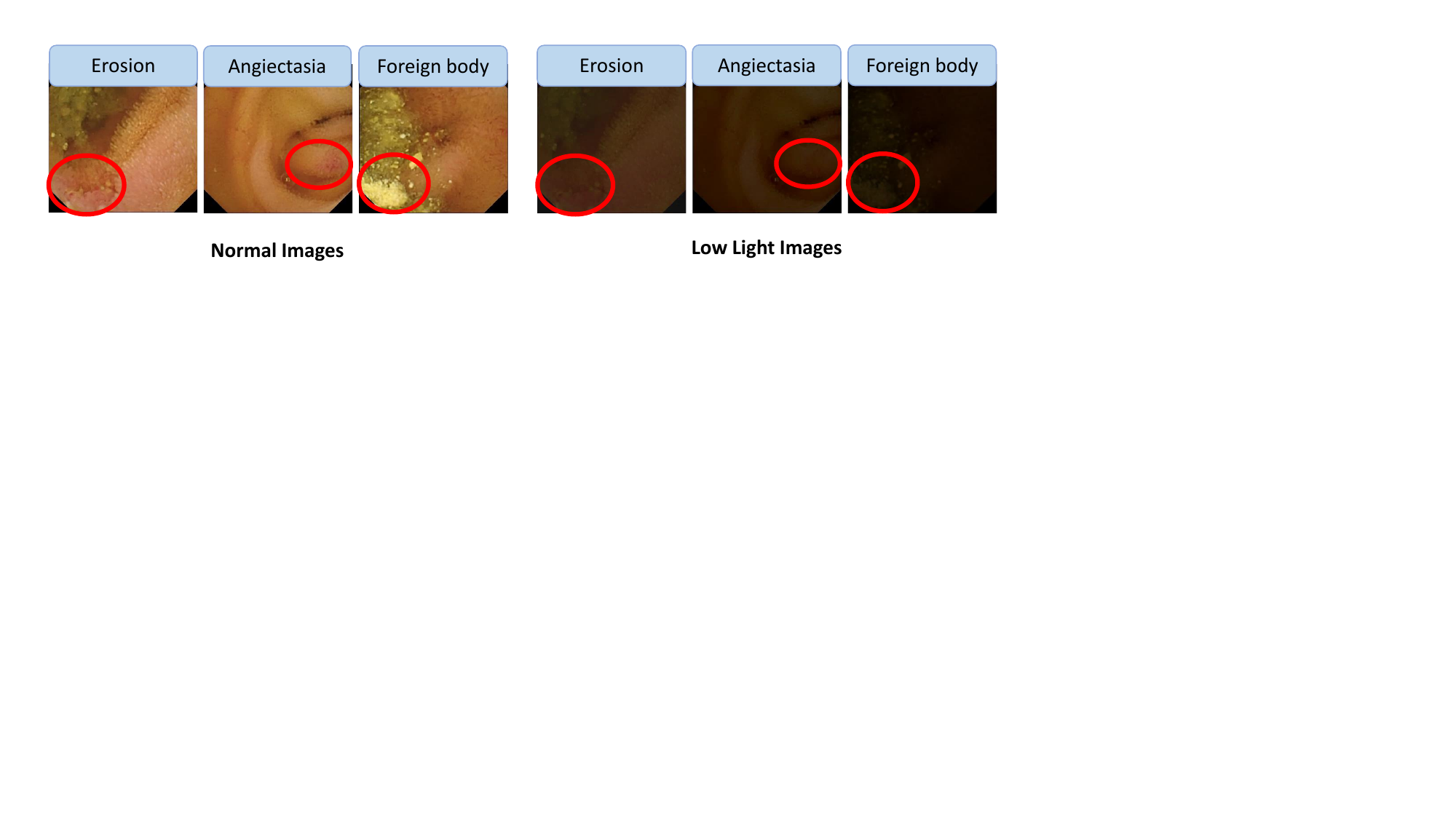}
    \caption{Comparison of normal images with low-light images. Obvious lesions are visible on normal images, but the same lesions can hardly be distinguished by human eyes in the corresponding low-light images.
    }
    \label{fig:compare}
\end{figure*}

Currently, the golden standard of gastrointestinal (GI) examination is endoscope screening, which can provide direct vision signals for diagnosis and analysis. Benefiting from its characteristics of being non-invasive, painless, and low physical burden, wireless capsule endoscopy (WCE) has the potential to overcome the shortcomings of conventional endoscopy~\cite{sliker2014flexible,zhang2022deep}. However, due to the anatomical complexity, insufficient illumination, and limited performance of the camera, low-quality images may hinder the diagnosis process~\cite{che2023image}. Blood vessels and lesions with minor color changes in the early stages can be hard to be screened out~\cite{long2018adaptive,bai2022transformer}. Fig.~\ref{fig:compare} shows WCE images with low illumination and contrast. The disease features clearly visible in the normal image become challenging to be found in the low-light images. Therefore, it is necessary to develop a low-light image enhancement framework for WCE to assist clinical diagnosis.
 
Many traditional algorithms (e.g., intensity transformation~\cite{guo2016lime}, histogram equalization~\cite{liu2016contrast}, and Retinex theory~\cite{li2018structure}) have been proposed for low-light image enhancement (LLIE). For WCE, Long~\textit{et al.}~\cite{long2018adaptive} discussed adaptive fraction-power transformation for image enhancement. However, traditional methods usually require an ideal assumption or an effective prior, limiting their wider applications. Deep learning (DL) provides novel avenues to solve LLIE problems~\cite{lore2017llnet, jiang2021enlightengan, guo2020zero}. Some DL-based LLIE schemes for medical endoscopy have been proposed~\cite{gomez2019low, ma2020cycle}. Gomez~\textit{et al.}~\cite{gomez2019low} offered a solution for laryngoscope low-light enhancement, and Ma~\textit{et al.}~\cite{ma2020cycle} proposed a medical image enhancement model with unpaired training data. 

Recently, denoising diffusion probabilistic model (DDPM)~\cite{ho2020DDPM} is the most popular topic in image generation, and has achieved success in various applications. Due to its unique regression process, DDPM has a stable training process and excellent output results, but also suffers from its expensive sampling procedure and lack of low-dimensional representation~\cite{pandey2022diffusevae}. It has been proved that DDPM can be combined with other existing DL techniques to speed up the sampling process~\cite{pandey2022diffusevae}. In our work, we introduce the reverse diffusion process of DDPM into our end-to-end LLIE process, which can preserve image details without introducing excessive computational costs. Our contributions to this work can be summarized as three-fold: 
\begin{itemize}
    \item [--]We design a \textbf{L}ow-\textbf{L}ight image enhancement framework for \textbf{Caps}ule endoscopy (\textbf{LLCaps}). Subsequent to the feature learning and preliminary shallow image reconstruction by the convolutional neural network (CNN), the reverse diffusion process is employed to further promote image reconstruction, preserve image details, and close in the optimization target.
    \item [--]Our proposed curved wavelet attention (CWA) block can efficiently extract high-frequency detail features via wavelet transform, and conduct local representation learning with the curved attention layer. 
    \item [--]Extensive experiments on two publicly accessible datasets demonstrate the excellent performance of our proposed model and components. The high-level lesion segmentation tasks further show the potential power of LLCaps on clinical applications.
\end{itemize}

\section{Methodology}
\subsection{Preliminaries} 

\subsubsection{Multi-scale Residual Block} Multi-scale Residual Block (MSRB)~\cite{li2018multi} constructs a multi-scale neuronal receptive field, which allows the network to learn multi-scale spatial information in the same layer. Therefore, the network can acquire contextual information from the low-resolution features while preserving high-resolution representations. We establish our CNN branch with six stacked multi-scale residual blocks (MSRB), and every two MSRBs are followed by a 2D convolutional layer (Conv2D). Besides, each MSRB shall require feature learning and the multi-scale feature aggregation module. Specifically, we propose our curved wavelet attention (CWA) module to conduct multi-scale feature learning, and employ the selective kernel feature fusion (SKFF)~\cite{zamir2020MIRNetv1} to combine multi-scale features, as shown in Fig.~\ref{fig:mainframe}~(a). 

\subsubsection{Denoising Diffusion Probabilistic Models} Denoising Diffusion Probabilistic Models (DDPMs)~\cite{ho2020DDPM} can be summarised as a model consisting of a forward noise addition $q{(i_{1:T}|i_{0})}$ and a reverse denoising process $p{(i_{0:T})}$, which are both parameterized Markov chains. The forward diffusion process gradually adds noise to the input image until the original input is destroyed. Correspondingly, the reverse process uses the neural network to model the Gaussian distribution and achieves image generation through gradual sampling and denoising.

\subsection{Proposed Methodology}

\begin{figure*}[t]
    \centering
    \includegraphics[width=1\linewidth, trim=0 20 10 0]{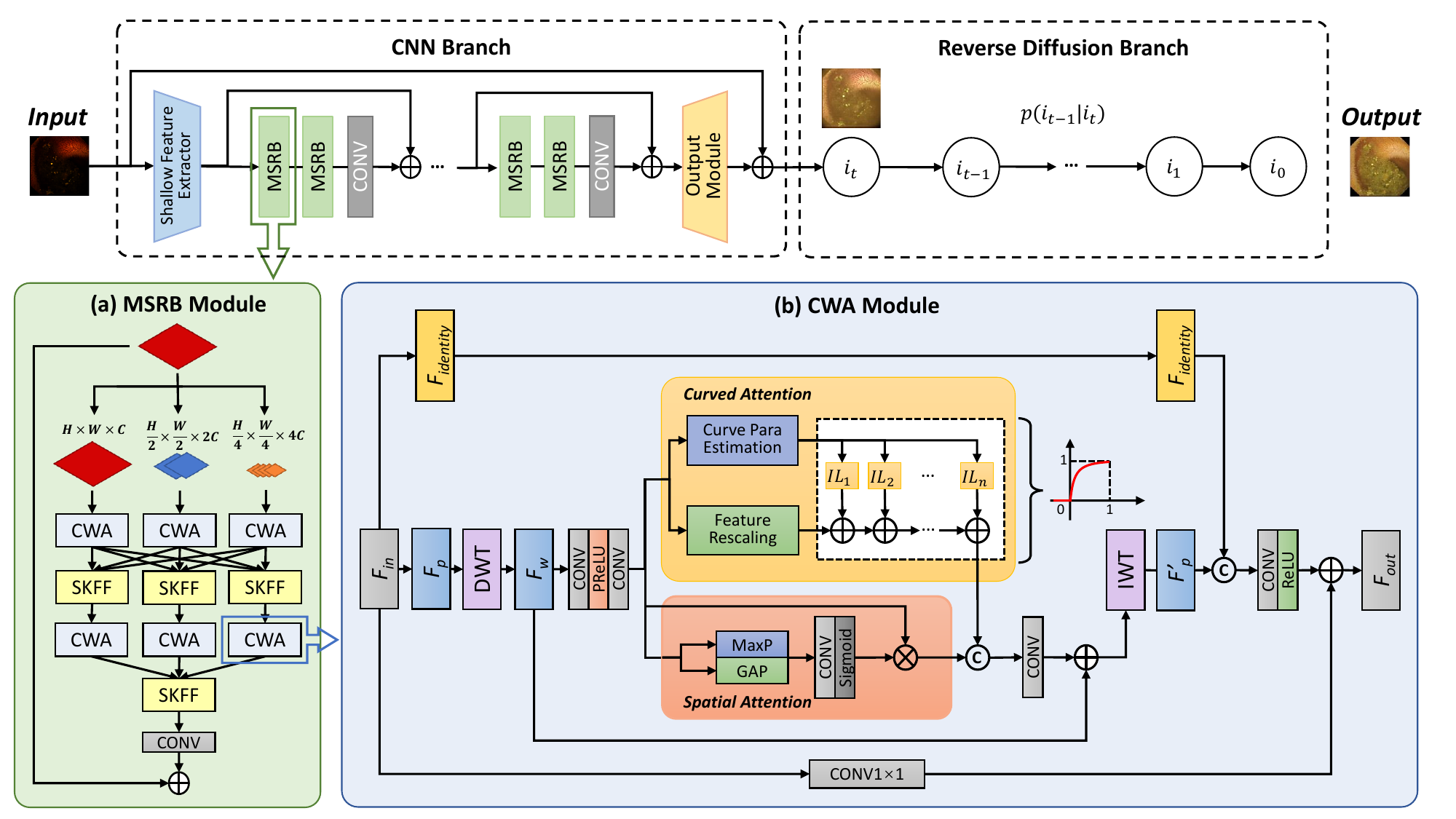}
    \caption{The overview of our proposed LLCaps. The CNN branch shall extract the shallow image output while the DDPM branch further optimizes the image via Markov chain inference. (a) represents the multi-scale residual block (MSRB), which allows the model to learn representation on different resolutions. (b) denotes our curved wavelet attention (CWA) block for attention learning and feature restoration. In (b), DWT and IWT denote discrete wavelet transform and inverse wavelet transform, respectively. The PReLU with two convolutional layers constructs the feature selector. `MaxP' denotes max pooling, and `GAP' means global average pooling.
    }
    \label{fig:mainframe}
\end{figure*}

\subsubsection{Curved Wavelet Attention}
Curved Wavelet Attention (CWA) block is the core component of our CNN branch, which is constructed via a curved dual attention mechanism and wavelet transform, as shown in Fig.~\ref{fig:mainframe}~(b). Firstly, the input feature map $F_{in}$ is divided into identity feature $F_{identity}$ and processing feature $F_{p}$. Medical LLIE shall require high image details. In this case, we transform the $F_{p}$ into wavelet domain $F_{w}$ to extract high-frequency detail information based on discrete wavelet transform. $F_{w}$ is then propagated through the feature selector and dual attention module for deep representation learning. Finally, we conduct reverse wavelet transform (IWT) to get ${F'}_{p}$, and concatenate it with $F_{identity}$ before the final output convolution layer.

We construct our curved dual attention module with parallel spatial and curved attention blocks. The spatial attention (SA) layer exploits the inter-spatial dependencies of convolutional features~\cite{zamir2020MIRNetv1}. The SA layer performs the global average pooling and max pooling on input features respectively, and concatenates the output ${F}_{w(mean)}$ and ${F}_{w(max)}$ to get ${F_{cat}}$. Then the feature map will be dimensionally reduced and passed through the activation function.

However, literature~\cite{guo2020zero, zhou2022lednet} has discussed the problem of local illumination in LLIE. If we simply use a global computing method such as the SA layer, the model may not be able to effectively understand the local illumination/lack of illumination. Therefore, in order to compensate for the SA layer, we design the Curved Attention (CurveA) layer, which is used to model the high-order curve of the input features. Let $\rm{IL}_{n(c)}$ denote the curve function, $c$ denote the feature location coordinates, and $Curve_{(n-1)}$ denote the pixel-wise curve parameter, we can obtain the curve estimation equation as: 
\begin{equation}
    \begin{aligned}
    \frac{\rm{IL}_{n(c)}}{\rm{IL}_{n-1(c)}}= Curve_{n-1}(1-\rm{IL}_{n-1(c)})
    \end{aligned}
    \label{equ:2}
\end{equation}
The detailed CurveA layer is presented in the top of Fig.~\ref{fig:mainframe}~(b), and the Equ.~(\ref{equ:2}) is related to the white area. The Curve Parameter Estimation module consists of a Sigmoid activation and several Conv2D layers, and shall estimate the pixel-wise curve parameter at each order. The Feature Rescaling module will rescale the input feature into [0, 1] to learn the concave down curves. By applying the CurveA layer to the channels of the feature map, the CWA block can better estimate local areas with different illumination.

\subsubsection{Reverse Diffusion Process}
Some works~\cite{wu2023medsegdiff, pandey2022diffusevae} have discussed combining diffusion models with other DL-based methods to reduce training costs and be used for downstream applications. In our work, We combine the reverse diffusion process of DDPM in a simple and ingenious way, and use it to optimize the shallow output by the CNN branch. Various experiments shall prove the effectiveness of our design in improving image quality and assisting clinical applications.

In our formulation, we assume that $i_0$ is the learning target $Y^*$ and $i_T$ is the output shallow image from the CNN branch. Therefore, we only need to engage the reverse process in our LLIE task. The reverse process is modeled using a Markov chain:
\begin{equation}
\begin{aligned}
& p_\theta\left(i_{0: T}\right)=p\left(i_T\right) \prod_{t=1}^T p_\theta\left(i_{t-1} \mid i_t\right) \\
& p_\theta\left(i_{t-1} \mid i_t\right)=\mathcal{N}\left(i_{t-1} ; \boldsymbol{\mu}_\theta\left(i_t, t\right), \mathbf{\Sigma}_\theta\left(i_t, t\right)\right)
\end{aligned}
\end{equation}
$p_\theta\left(i_{t-1} \mid i_t\right)$ are parameterized Gaussian distributions whose mean $\boldsymbol{\mu}_\theta\left(i_t, t\right)$ and variance $\mathbf{\Sigma}_\theta\left(i_t, t\right)$ are given by the trained network. Meanwhile, we simplify the network and directly include the reverse diffusion process in the end-to-end training of the entire network. Shallow output is therefore optimized by the reverse diffusion branch to get the predicted image $Y$. We further simplify the optimization function and only employ a pixel-level loss on the final output image, which also improves the training and convergence efficiency. 

\subsubsection{Overall Network Architecture}
\label{sec:overall}
An overview of our framework can be found in Fig.~\ref{fig:mainframe}. Our LLCaps contains a CNN branch (including a shallow feature extractor (SFE), multi-scale residual blocks (MSRBs), an output module (OPM)), and the reverse diffusion process. The SFE is a Conv2D layer that maps the input image into the high-dimensional feature representation $F_{SFE}\in\mathbb R^{C \times W \times H}$~\cite{xiao2021early}. Stacked MSRBs shall conduct deep feature extraction and learning. OPM is a Conv2D layer that recovers the feature space into image pixels. A residual connection is employed here to optimize the end-to-end training and converge process. Hence, given a low-light image $x\in\mathbb R^{3 \times W \times H}$, where $W$ and $H$ represent the width and height, the CNN branch can be formulated as:
\begin{equation}
\begin{aligned}
  & F_{SFE}=H_{SFE}(x) \\
  & F_{MSRBs}=H_{MSRBs}(F_{SFE}),\\
  & F_{OPM}=H_{OPM}({F_{MB}})+x  
\end{aligned}
\end{equation}

The shallow output $F_{OPM}\in\mathbb R^{3 \times W \times H}$ shall further be propagated through the reverse diffusion process and achieve the final enhanced image $Y\in\mathbb R^{3 \times W \times H}$. The whole network is constructed in an end-to-end mode and optimized by Charbonnier loss~\cite{charbonnier1994two}. The $\varepsilon$ is set to $10^{-3}$ empirically.
\begin{equation} 
\mathcal{L}\left(x, x^*\right)=\sqrt{\left\|Y - Y^*\right\|^2+\varepsilon^2}  
\end{equation}
in which $Y$ and $Y^*$ denote the input and ground truth images, respectively.

\section{Experiments}
\label{sec:exper}

\subsection{Dataset}
\label{sec:dataset}

We conduct our experiments on two publicly accessible WCE datasets, the Kvasir-Capsule~\cite{smedsrud2021kvasir} and the Red Lesion Endoscopy (RLE) dataset~\cite{coelho2018RLE}.

\textbf{Kvasir-Capsule dataset}~\cite{smedsrud2021kvasir} is a WCE classification dataset with three anatomy classes and eleven luminal finding classes. By following~\cite{chen2022dynamic}, we randomly select 2400 images from the Kvasir-Capsule dataset, of which 2000 are used for training and 400 for testing. To create low-light images, we adopt random Gamma correction and illumination reduction following~\cite{lore2017llnet,li2021low}. Furthermore, to evaluate the performance on real data, we add an external validation on 100 real images selected from the Kvasir-Capsule dataset. These images are with low brightness and are not included in our original experiments.

\textbf{Red Lesion Endoscopy dataset}~\cite{coelho2018RLE} (RLE) is a WCE dataset for red lesion segmentation tasks (e.g., angioectasias, angiodysplasias, and bleeding). We randomly choose 1283 images, of which 946 images are used for training and 337 for testing. We adopt the same method in the Kvasir-Capsule dataset to generate low-light images. Furthermore, we conduct a segmentation task on the RLE test set to investigate the effectiveness of the LLIE models in clinical applications.

\subsection{Implementation Details}
\label{sec:implementation}

\begin{figure*}[!t]
    \centering
    \includegraphics[width=0.95\linewidth, trim=60 60 10 0]{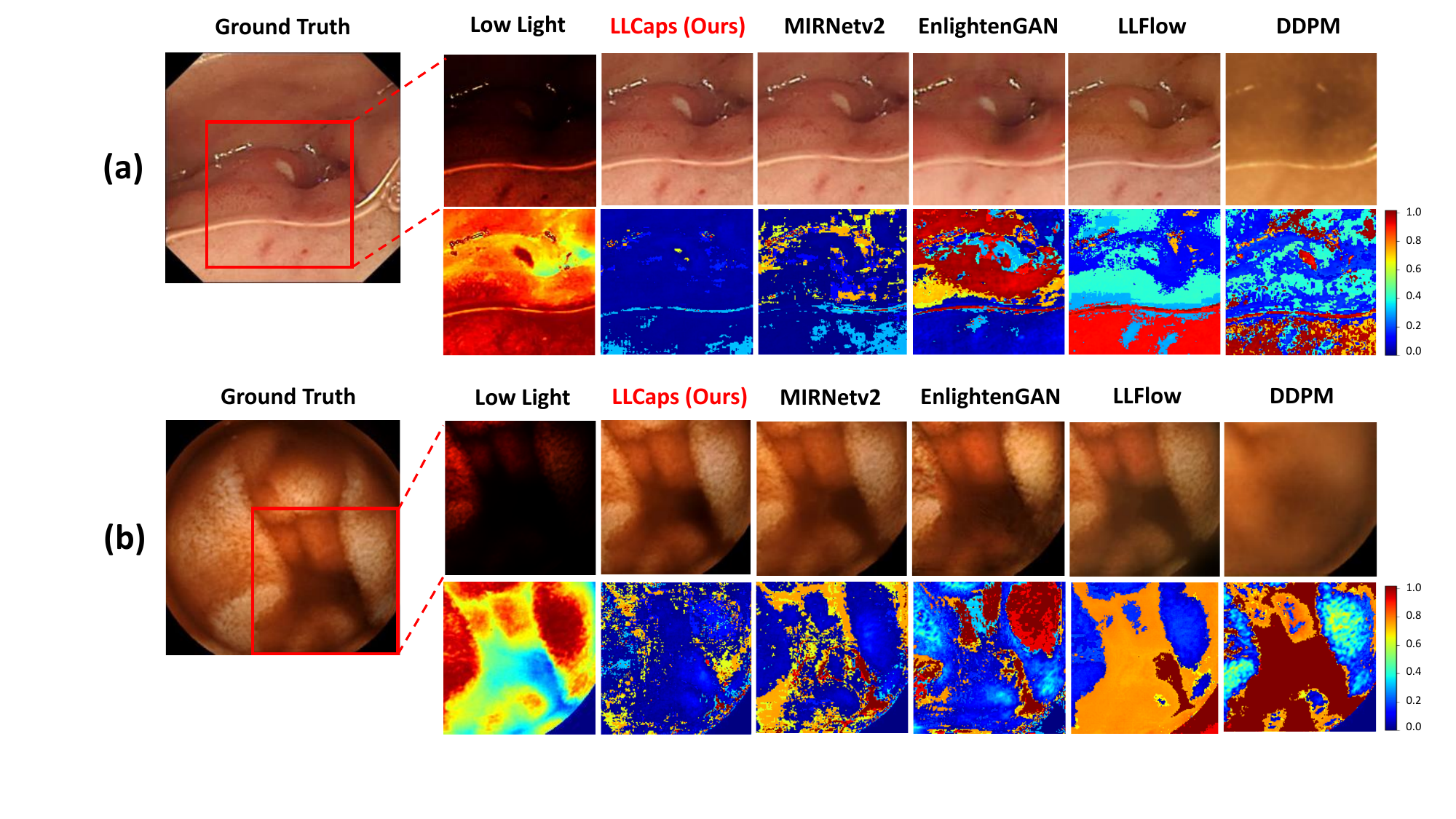}
    \caption{The quantitative results for LLCaps compared with SOTA approaches on (a) Kvasir-Capsule dataset~\cite{smedsrud2021kvasir} and (b) RLE dataset~\cite{coelho2018RLE}. The first row visualizes the enhanced images from different LLIE approaches, and the second row contains the reconstruction error heat maps. The blue and red represent low and high error, respectively.
    }
    \label{fig:IMG_visual}
\end{figure*}

We compare the performance of our LLCaps against the following state-of-the-art (SOTA) LLIE methodologies: LIME~\cite{guo2016lime}, DUAL~\cite{zhang2019dual}, Zero-DCE~\cite{guo2020zero}, EnlightenGAN~\cite{jiang2021enlightengan}, LLFlow~\cite{wang2022llflow}, HWMNet~\cite{fan2022half}, MIRNet~\cite{zamir2020MIRNetv1}, SNR-Aware~\cite{xu2022snr}, StillGAN~\cite{ma2021structure}, MIRNetv2~\cite{zamir2022MIRNetv2}, and DDPM~\cite{ho2020DDPM}. Our models are trained using Adam optimizer for $200$ epochs with a batch size of $4$ and a learning rate of $1 \times 10^{-4}$. For evaluation, we adopt three commonly used image quality assessment metrics: Peak Signal-to-Noise Ratio (PSNR)~\cite{huynh2008scope}, Structural Similarity Index (SSIM)~\cite{wang2004image}, and Learned Perceptual Image Patch Similarity (LPIPS)~\cite{zhang2018unreasonable}. 
For the external validation set, we evaluate with no-reference metrics LPIPS~\cite{zhang2018unreasonable} and Perception-based Image Quality Evaluator (PIQE)~\cite{venkatanath2015blind} due to the lack of ground truth images. To verify the usefulness of the LLIE methods for downstream medical tasks, we conduct red lesion segmentation on the RLE test set and evaluate the performance via mean Intersection over Union (mIoU), Dice similarity coefficient (Dice), and Hausdorff Distance (HD). We train UNet~\cite{ronneberger2015u} using Adam optimizer for $20$ epochs. The batch size and learning rate are set to $4$ and $1 \times 10^{-4}$, respectively. All experiments are implemented by Python PyTorch and conducted on NVIDIA RTX 3090 GPU. Results are the average of 3-fold cross-validation.

\subsection{Results}
\label{sec:results}

We compare the performance of our LLCaps to the existing approaches, as demonstrated in Table~\ref{tab:1} and Fig.~\ref{fig:IMG_visual} quantitatively and qualitatively. Compared with other methods, our proposed method achieves the best performance among all metrics. Specifically, our method surpasses MIRNetv2~\cite{zamir2022MIRNetv2} by 3.57 dB for the Kvasir-Capsule dataset and 0.33 dB for the RLE dataset. The SSIM of our method has improved to 96.34\% in the Kvasir-Capsule dataset and 93.34\% in the RLE dataset. Besides that, our method also performs the best in the no-reference metric LPIPS. The qualitative results of the comparison methods and our method on the Kvasir-Capsule and RLE datasets are visualized in Fig.~\ref{fig:IMG_visual} with the corresponding heat maps. Firstly, we can see that directly performing LLIE training on DDPM~\cite{ho2020DDPM} cannot obtain good image restoration, and the original structures of the DDPM images are largely damaged. EnlightenGAN~\cite{jiang2021enlightengan} also does not perform satisfactorily in structure restoration. Our method successfully surpasses LLFlow~\cite{wang2022llflow} and MIRNetv2~\cite{zamir2022MIRNetv2} in illumination restoration. The error heat maps further reflect the superior performance of our method in recovering the illumination and structure from low-light images. Moreover, our solution yields the best on the real low-light dataset during the external validation, proving the superior performance of our solution in real-world applications.

\begin{table}[t]
	\caption{
        Image quality comparison with existing methods on Kvasir-Capsule~\cite{smedsrud2021kvasir} and RLE dataset~\cite{coelho2018RLE}. The `External Val' denotes the external validation experiment conducted on 100 selected real low-light images from the Kvasir-Capsule dataset~\cite{smedsrud2021kvasir}. The red lesion segmentation experiment is also conducted on RLE test set~\cite{coelho2018RLE}.
	}
 	\centering
	\label{tab:1}  
 \resizebox{\textwidth}{!}{	
\begin{tabular}{c|ccc|ccc|cc|ccc}
\hline

\multirow{2}{*}{Models} & \multicolumn{3}{c|}{Kvasir-Capsule}          & \multicolumn{3}{c|}{RLE} & \multicolumn{2}{c|}{External Val} & \multicolumn{3}{c}{RLE Segmentation}  \\ \cline{2-12} 
& PSNR $\uparrow$ & SSIM $\uparrow$  & LPIPS $\downarrow$ & PSNR $\uparrow$ & SSIM $\uparrow$  & LPIPS $\downarrow$ & LPIPS $\downarrow$ & PIQE $\downarrow$ & mIoU $\uparrow$ & Dice $\uparrow$ & HD $\downarrow$ \\ \hline
LIME~\cite{guo2016lime}                   & 12.07 & 29.66 & 0.4401 & 14.21 & 15.93 & 0.5144 & 0.3498 & 26.41 & 60.19 & 78.42 & 56.20 \\
DUAL~\cite{zhang2019dual}                 & 11.61 & 29.01 & 0.4532 & 14.64 & 16.11 & 0.4903 & 0.3305 & 25.47 & 61.89 & 78.15 & 55.70 \\
Zero-DCE~\cite{guo2020zero}               & 14.03 & 46.31 & 0.4917 & 14.86 & 34.18 & 0.4519 & 0.6723 & 21.47 & 54.77 & 71.46 & 56.24 \\
EnlightenGAN~\cite{jiang2021enlightengan} & 27.15 & 85.03 & 0.1769 & 23.65 & 80.51 & 0.1864 & 0.4796 & 34.75 & 61.97 & 74.15 & 54.89 \\
LLFlow~\cite{wang2022llflow}              & 29.69 & 92.57 & 0.0774 & 25.93 & 85.19 & 0.1340 & 0.3712 & 35.67 & 61.06 & \textbf{78.55} & 60.04  \\
HWMNet~\cite{fan2022half}                 & 27.62 & 92.09 & 0.1507 & 21.81 & 76.11 & 0.3624 & 0.5089 & 35.37 & 56.48 & 74.17 & 59.90 \\
MIRNet~\cite{zamir2020MIRNetv1}           & 31.23 & 95.77 & 0.0436 & 25.77 & 86.94 & 0.1519 & 0.3485 & 34.28 & 59.84 & 78.32 & 63.10 \\
StillGAN~\cite{ma2021structure}           & 28.28 & 91.30 & 0.1302 & 26.38 & 83.33 & 0.1860 & 0.3095 & 38.10 & 58.32 & 71.56 & 55.02 \\
SNR-Aware~\cite{xu2022snr}                & 30.32 & 94.92 & 0.0521 & 27.73 & 88.44 & 0.1094 & 0.3992 & 26.82 & 58.95 & 70.26 & 57.73 \\
MIRNetv2~\cite{zamir2022MIRNetv2}         & 31.67 & 95.22 & 0.0486 & 32.85 & 92.69 & 0.0781 & 0.3341 & 41.24 & 63.14 & 75.07 & 53.71 \\
DDPM~\cite{ho2020DDPM}                    & 25.17 & 73.16 & 0.4098 & 22.97 & 70.31 & 0.4198 & 0.5069 & 43.64 & 54.09 & 75.10 & 67.54 \\
\textbf{LLCaps (Ours)} & \textbf{35.24} & \textbf{96.34} & \textbf{0.0374} & \textbf{33.18} & \textbf{93.34} & \textbf{0.0721} & \textbf{0.3082} & \textbf{20.67} & \textbf{66.47} & 78.47 & \textbf{44.37} \\ \hline
\end{tabular}}
\end{table}

Furthermore, a downstream red lesion segmentation task is conducted to investigate the usefulness of our LLCaps on clinical applications.
As illustrated in Table~\ref{tab:1}, LLCaps achieve the best lesion segmentation results, manifesting the superior performance of our LLCaps model in lesion segmentation. Additionally, LLCaps surpasses all SOTA methods in HD, showing LLCaps images perform perfectly in processing the segmentation boundaries, suggesting that our method possesses better image reconstruction and edge retention ability.

Besides, an ablation study is conducted on the Kvasir-Capsule dataset to demonstrate the effectiveness of our design and network components, as shown in Table~\ref{tab:ablation}. To observe and compare the performance changes, we try to (i) remove the wavelet transform in CWA blocks, (ii) degenerate the curved attention (CurveA) layer in CWA block to a simple channel attention layer~\cite{zamir2020MIRNetv1}, and (iii) remove the reverse diffusion branch. Experimental results demonstrate that the absence of any component shall cause great performance degradation. The significant improvement in quantitative metrics is a further testament to the effectiveness of our design for each component.

\begin{table}[t]
\centering
	\caption{
        Ablation experiments of our LLCaps on the Kvasir-Capsule Dataset~\cite{smedsrud2021kvasir}. In order to observe the performance changes, we (i) remove the wavelet transform, (ii)~degenerate the CurveA layer, and (iii) remove the reverse diffusion branch.
	}
 	\centering
        \resizebox{.68\textwidth}{!}{
 \label{tab:ablation}
\begin{tabular}{c|c|c|p{1.5cm}<{\centering} p{1.5cm}<{\centering} p{1.5cm}<{\centering}}
\hline
\multicolumn{1}{c|}{\makecell[c]{Wavelet\\Transform }} & \multicolumn{1}{c|}{\makecell[c]{Curve\\Attention}} & \multicolumn{1}{c|}{\makecell[c]{Reverse\\Diffusion}} & PSNR $\uparrow$ & SSIM $\uparrow$  & LPIPS $\downarrow$  \\ \hline
\XSolidBrush      &\XSolidBrush      & \multicolumn{1}{c|}{\XSolidBrush}         & 31.12 & 94.96 & 0.0793  \\
\Checkmark      &\XSolidBrush      & \multicolumn{1}{c|}{\XSolidBrush}         & 32.78 & 96.26 & 0.0394  \\
\XSolidBrush       & \Checkmark     & \multicolumn{1}{c|}{\XSolidBrush}         & 32.08 & 96.27 & 0.0415 \\
 \XSolidBrush      &\XSolidBrush      & \multicolumn{1}{c|}{\Checkmark}         & 33.10 & 94.53 & 0.0709 \\
\Checkmark       & \Checkmark     & \multicolumn{1}{c|}{\XSolidBrush}         & 33.92 & 96.20 & 0.0381 \\
\Checkmark      &\XSolidBrush      & \multicolumn{1}{c|}{\Checkmark}         & 34.07 & 95.61 & 0.0518  \\
\XSolidBrush       & \Checkmark    & \multicolumn{1}{c|}{\Checkmark}         & 33.41 & 95.03 & 0.0579 \\
\Checkmark     & \Checkmark     & \multicolumn{1}{c|}{\Checkmark}         & \textbf{35.24} & \textbf{96.34} & \textbf{0.0374}  \\ \hline
\end{tabular}}
\end{table}

\section{Conclusion}
\label{sec:conclusion}
We present LLCaps, an end-to-end capsule endoscopy LLIE framework with multi-scale CNN and reverse diffusion process. The CNN branch is constructed by stacked MSRB modules, in which the core CWA block extracts high-frequency detail information through wavelet transform, and learns the local representation of the image via the Curved Attention layer. The reverse diffusion process further optimizes the shallow output, achieving the closest approximation to the real image. Comparison and ablation studies prove that our method and design bring about superior performance improvement in image quality. Further medical image segmentation experiments demonstrate the reliability of our method in clinical applications. Potential future works include extending our model to various medical scenarios (e.g., surgical robotics, endoscopic navigation, augmented reality for surgery) and clinical deep learning model deployment.

\subsubsection{Acknowledgements.}
This work was supported by Hong Kong RGC CRF C4063-18G,  CRF C4026-21GF, RIF R4020-22, GRF 14216022, GRF 14211420,  NSFC/RGC JRS N\_CUHK420/22; Shenzhen-Hong Kong-Macau Technology Research Programme (Type C 202108233000303); GBABF \#2021B1515120035.

%
%
%
%
\bibliography{reference}{}
\bibliographystyle{splncs04}

\newpage
\section*{Supplementary Materials for ``LLCaps: Learning to Illuminate Low-Light Capsule Endoscopy with Curved Wavelet Attention and Reverse Diffusion''}

\begin{figure}
\centering
\includegraphics[width=\textwidth, trim=20 50 0 0]{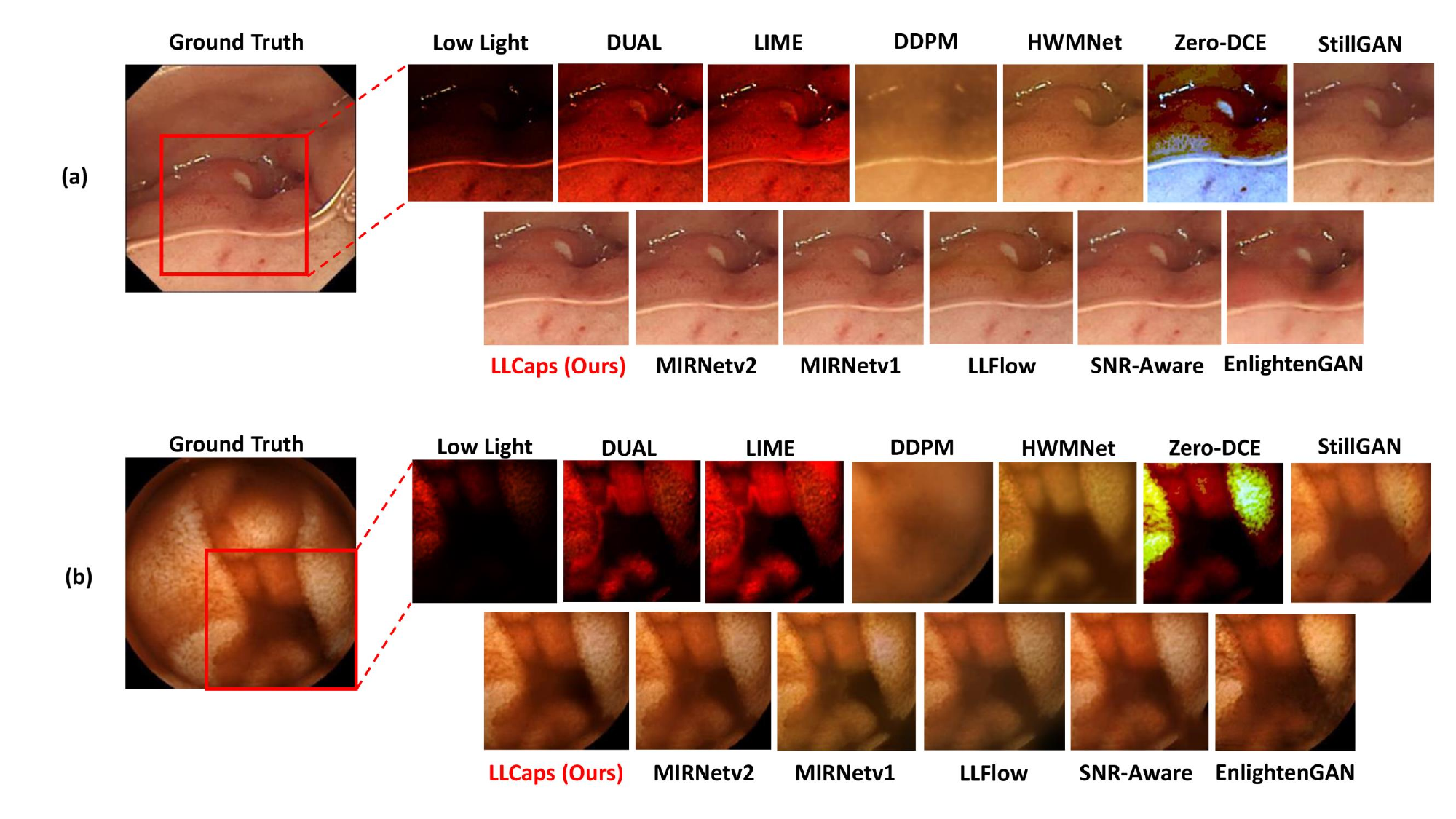}
\caption{The full visualization results for LLCaps compared with SOTA approaches on (a) Kvasir-Capsule dataset and (b) RLE dataset.} \label{figa1}
\end{figure}

\begin{figure}
\centering
\includegraphics[width=\textwidth, trim=0 20 0 100]{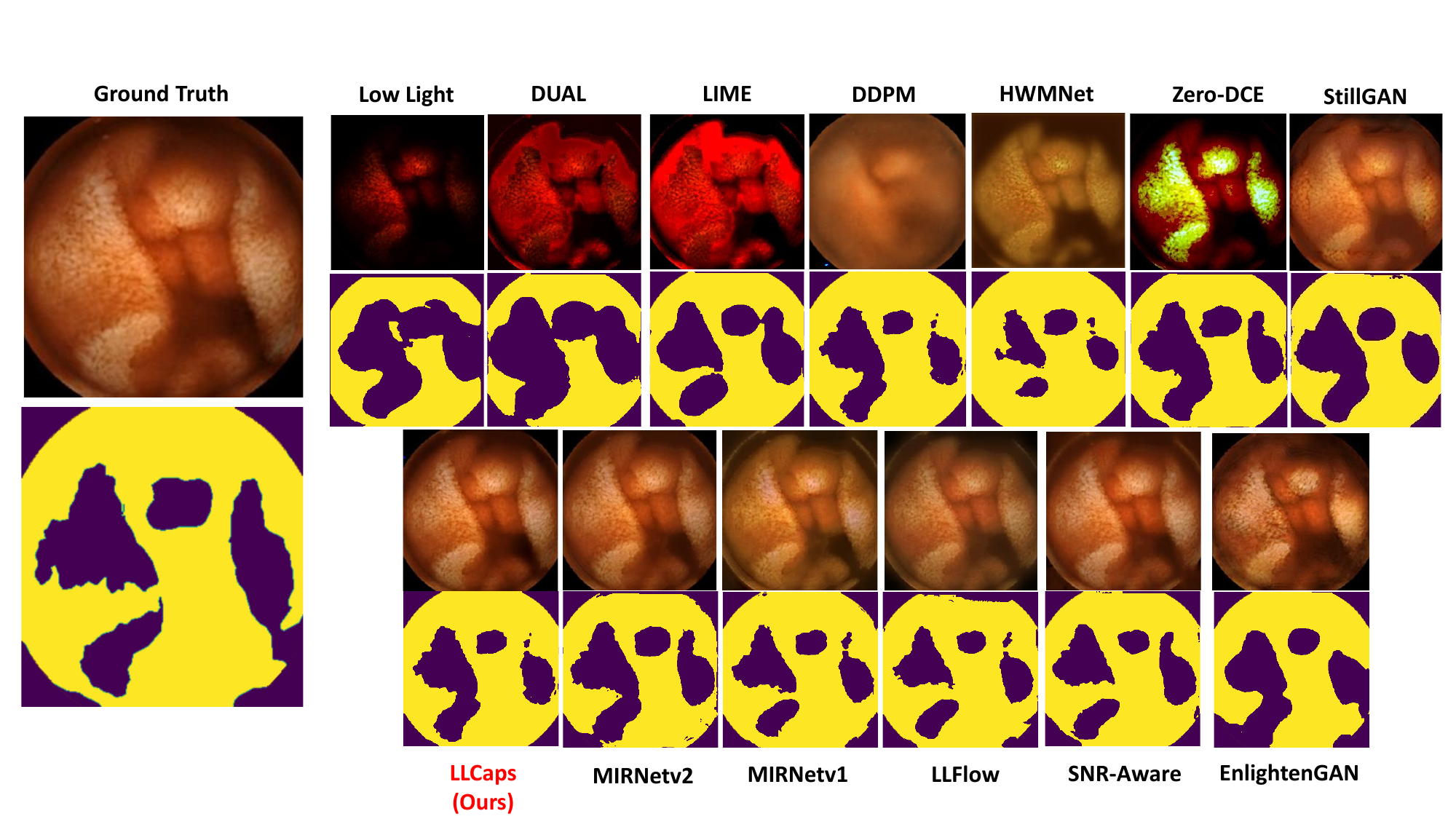}

\caption{Visualization of the red lesion segmentation comparison experiments on the RLE dataset.}
\end{figure}

\begin{table}
	\caption{
Ablation study of the frequency analysis on the CWA block using average gradient. Higher mean and variance of AG denote richer details. When removing the wavelet transform or the CWA block, the mean and var drop greatly, showing the effectiveness of our proposed CWA block in extracting features.
	}
 	\centering
	\label{tab:1}  
\begin{tabular}{l|c|c}
\hline

Average Gradient & Mean $\times 10^5 \; \uparrow$ & Variance $\times 10^{10} \; \uparrow$ \\ \hline
DDPM $[9]$ & 2.19 & 0.50 \\
MIRNetv2 $[30]$ & 3.51 & 1.86 \\
LLCaps w/o CWA & 3.54 & 0.99 \\
LLCaps w/o Curved Attention & 3.85 & 2.01 \\
LLCaps w/o Wavelet Transform & 3.55 & 1.93 \\
LLCaps & 3.87 & 2.09 \\
GT & 3.95 & 2.19 \\ \hline
\end{tabular}
\end{table}



\end{document}